\documentclass[final,1p,times,twocolumn]{elsarticle}
\usepackage{graphicx,amssymb}
\journal{Physics Letters B}
\begin{document}
\begin{frontmatter}

\title{Updated constraints on spatial variations of the fine-structure constant}
\author[inst1,inst2]{A. M. M. Pinho}\ead{Ana.Pinho@astro.up.pt}
\author[inst1,inst3]{C. J. A. P. Martins\corref{cor1}}\ead{Carlos.Martins@astro.up.pt}
\address[inst1]{Centro de Astrof\'{\i}sica, Universidade do Porto, Rua das Estrelas, 4150-762 Porto, Portugal}
\address[inst2]{Faculdade de Ci\^encias, Universidade do Porto, Rua do Campo Alegre 687, 4169-007 Porto, Portugal}
\address[inst3]{Instituto de Astrof\'{\i}sica e Ci\^encias do Espa\c co, CAUP, Rua das Estrelas, 4150-762 Porto, Portugal}
\cortext[cor1]{Corresponding author}

\begin{abstract}
Recent work by Webb {\it et al.} has provided indications of spatial variations of the fine-structure constant, $\alpha$, at a level of a few parts per million. Using a dataset of 293 archival measurements, they further show that a dipole provides a statistically good fit to the data, a result subsequently confirmed by other authors. Here we show that a more recent dataset of dedicated measurements further constrains these variations: although there are only 10 such measurements, their uncertainties are considerably smaller. We find that a dipolar variation is still a good fit to the combined dataset, but the amplitude of such a dipole must be somewhat smaller: $8.1\pm1.7$ ppm for the full dataset, versus $9.4\pm2.2$ ppm for the  Webb {\it et al.} data alone, both at the $68.3\%$ confidence level. Constraints on the direction on the sky of such a dipole are also significantly improved. On the other hand the data can't yet discriminate between a pure spatial dipole and one with an additional redshift dependence.
\end{abstract}


\end{frontmatter}

\section{Introduction}

Testing the stability of nature's fundamental couplings is among the most actively pursued topics in observational astrophysics \cite{Uzan}. In addition to the intrinsically fundamental nature of these tests, the measurements (whether they are detections of variations or null results) have deep consequences for cosmology and fundamental physics, an overview of which is provided in \cite{grg}.

A recent analysis by Webb {\it et al.}  of a large archival dataset has provided some evidence for spatial variations of the fine-structure constant, $\alpha$, at the level of a few parts per million (ppm) \cite{Webb,King}. The dataset includes a total of 293 measurements in the approximate redshift range $0.2<z.<4.2$, obtained with ESO's UVES spectrograph at the VLT and with the HIRES spectrograph at the Keck telescope. Both the analysis of Webb {\it et al.} and those of subsequent works \cite{tests1,tests2,tests3,tests4} find evidence for a spatial dipole in the measurements, at a statistical level of significance of more than four standard deviations.

Meanwhile some dedicated measurements of $\alpha$ (that is, those where the data was specifically taken for this purpose) has been obtained and further efforts in this direction are ongoing, such as those of the UVES Large Program for Testing Fundamental Physics \cite{LP1,LP3}. The number of currently available dedicated measurements is only a dozen or so, so they can't yet be used on their own to search for spatial variations. Nevertheless, these measurements have statistical and systematic uncertainties that are nominally smaller than those of the archival measurements. (Note that in a large sample such as that of Webb {\it et al.} the systematic uncertainties can be---and have been---estimated directly from the sample distribution, while this is not possible for individual measurements.) Here, therefore, we carry out a first joint analysis of the  Webb {\it et al.} and the more recent measurements, with the aim of ascertaining whether the evidence for the dipolar variation is preserved.

\section{Available data and parameterizations}

Previous studies of the spatial distribution of $\alpha$ measurements were restricted to the data of Webb {\it et al.} \cite{Webb}, which is a large dataset of archival data measurements. This dataset has been extensively described elsewhere (most notably in \cite{King}), and we refer the reader to these works for additional details. There have been recent suggestions that the level of systematics in these measurements may have been underestimated \cite{Whitmore}, but here we simply take the published values at face value, and calculate the total uncertainty for each measurement by adding in quadrature the statistical and systematic uncertainties.

In our analysis we will also consider this data on its own (to check that we recover previously published results) but, more importantly, we will for the first time combine it with the available, smaller and more recent, dataset of dedicated measurements listed in Table \ref{tablealpha}. This compilation includes the early results of the UVES Large Program for Testing Fundamental Physics \cite{LP1,LP3}, which is expected to be the one with a better control of possible systematics. The source of the data in this Table is also further discussed in \cite{Ferreira}.

We note that the first measurement listed on the table is the weighted average from measurements in 8 absorption systems in the redshift range $0.73<z<1.53$ along lines of sight that are widely separated on the sky (HE1104-1805A, HS1700+6416 and HS1946+7658) \cite{Songaila}; the authors only report this average and not the individual measurements. For this reason we listed the result in Table \ref{tablealpha} for completeness but naturally it won't be included in our analysis. Our more recent dataset therefore has 10 different measurements, all in the redshift range $1<z<2$.

\begin{table}
\begin{center}
\begin{tabular}{|c|c|c|c|c|}
\hline
 Object & z & ${ \Delta\alpha}/{\alpha}$ (ppm) & Spectrograph & Ref. \\
\hline\hline
3 sources & 1.08 & $4.3\pm3.4$ & HIRES & \protect\cite{Songaila} \\
\hline
HS1549$+$1919 & 1.14 & $-7.5\pm5.5$ & UVES/HIRES/HDS & \protect\cite{LP3} \\
\hline
HE0515$-$4414 & 1.15 & $-0.1\pm1.8$ & UVES & \protect\cite{alphaMolaro} \\
\hline
HE0515$-$4414 & 1.15 & $0.5\pm2.4$ & HARPS/UVES & \protect\cite{alphaChand} \\
\hline
HS1549$+$1919 & 1.34 & $-0.7\pm6.6$ & UVES/HIRES/HDS & \protect\cite{LP3} \\
\hline
HE0001$-$2340 & 1.58 & $-1.5\pm2.6$ &  UVES & \protect\cite{alphaAgafonova}\\
\hline
HE1104$-$1805A & 1.66 & $-4.7\pm5.3$ & HIRES & \protect\cite{Songaila} \\
\hline
HE2217$-$2818 & 1.69 & $1.3\pm2.6$ &  UVES & \protect\cite{LP1}\\
\hline
HS1946$+$7658 & 1.74 & $-7.9\pm6.2$ & HIRES & \protect\cite{Songaila} \\
\hline
HS1549$+$1919 & 1.80 & $-6.4\pm7.2$ & UVES/HIRES/HDS & \protect\cite{LP3} \\
\hline
Q1101$-$264 & 1.84 & $5.7\pm2.7$ &  UVES & \protect\cite{alphaMolaro}\\
\hline
\end{tabular}
\caption{\label{tablealpha}Recent dedicated measurements of $\alpha$. Listed are, respectively, the object along each line of sight, the redshift of the measurement, the measurement itself (in parts per million), the spectrograph(s), and the original reference. The recent UVES Large Program measurements are \cite{LP1,LP3}. The  quoted errors include both statistical and systematic uncertainties (to the extent that these were estimated in the original works), added in quadrature. The first measurement is the weighted average from 8 absorbers in the redshift range $0.73<z<1.53$ along the lines of sight of HE1104-1805A, HS1700+6416 and HS1946+7658, reported in \cite{Songaila} without the values for individual systems, and therefore won't be included in our analysis.}
\end{center}
\end{table}

We will fit this data to two different phenomenological parameterizations. The first is a pure spatial dipole for the relative variation of $\alpha$
\begin{equation}\label{puredipole}
\frac{\Delta\alpha}{\alpha}(A,\Psi)=A\cos{\Psi}\,,
\end{equation}
which depends on the orthodromic distance $\Psi$ to the north pole of the dipole (the locus of maximal positive variation) given by
\begin{equation}\label{ortho}
\cos{\Psi}=\sin{\theta_i}\sin{\theta_0}+\cos{\theta_i}\cos{\theta_0}\cos{(\phi_i-\phi_0)}\,,
\end{equation}
with $(\theta_i,\phi_i)$ being the Declination and Right Ascension of the i-th measurement and $(\theta_0,\phi_0)$ those of the north pole. These latter two coordinates, together with the overall amplitude $A$, are our free parameters. Such a parameterization has been considered in all previous analyses of the Webb {\it et al.} data \cite{Webb,King,tests1,tests2,tests3,tests4} and thus serves as a simple test of our analysis. We note that we do not consider an additional monopole term, both because there is no strong statistical preference for it in previous analyses \cite{Webb,King} and because physically such term would be understood as being due to the assumption of terrestrial isotopic abundances, in particular of Magnesium---we refer the interested reader to \cite{Monopole} for a detailed discussion of this point.

Additionally we will also consider a parameterization where there is an implicit time dependence in addition to the spatial variation. Previous analyses considered the case of a dependence on look-back time \cite{Webb,King}, but this has the disadvantage of requiring a specific assumption of a cosmological model, and moreover it's not clear how such a dependence would emerge from realistic varying $\alpha$ models. We will instead assume a logarithmic dependence on redshift
\begin{equation}\label{redshiftdipole}
\frac{\Delta\alpha}{\alpha}(A,z,\Psi)=A\, \ln{(1+z)}\, \cos{\Psi}\,;
\end{equation}
this has the advantage of not requiring any additional free parameters, but such dependencies are also typical of dilaton-type models \cite{Dilaton}. As in previous analyses, this parameterization is mainly considered as a means to assess the ability of the data to discriminate between models.

\section{Results}

We used standard likelihood techniques to fit the two parameterizations to our datasets. We considered grids of size $200^3$, for the Amplitude of the dipole and the Right Ascension and Declination of its north pole. We assumed a positive value of the amplitude and uniform priors on all three parameters. It is intuitively clear (but we have nevertheless explicitly checked it, as a further test of our analysis pipeline) that allowing also for negative values of the amplitude would lead to degenerate plots, with a specific amplitude and its negative equally likely and two opposite points on the sky also equally likely as the best-fit poles.

\begin{figure}
\begin{center}
\includegraphics[width=3in,keepaspectratio]{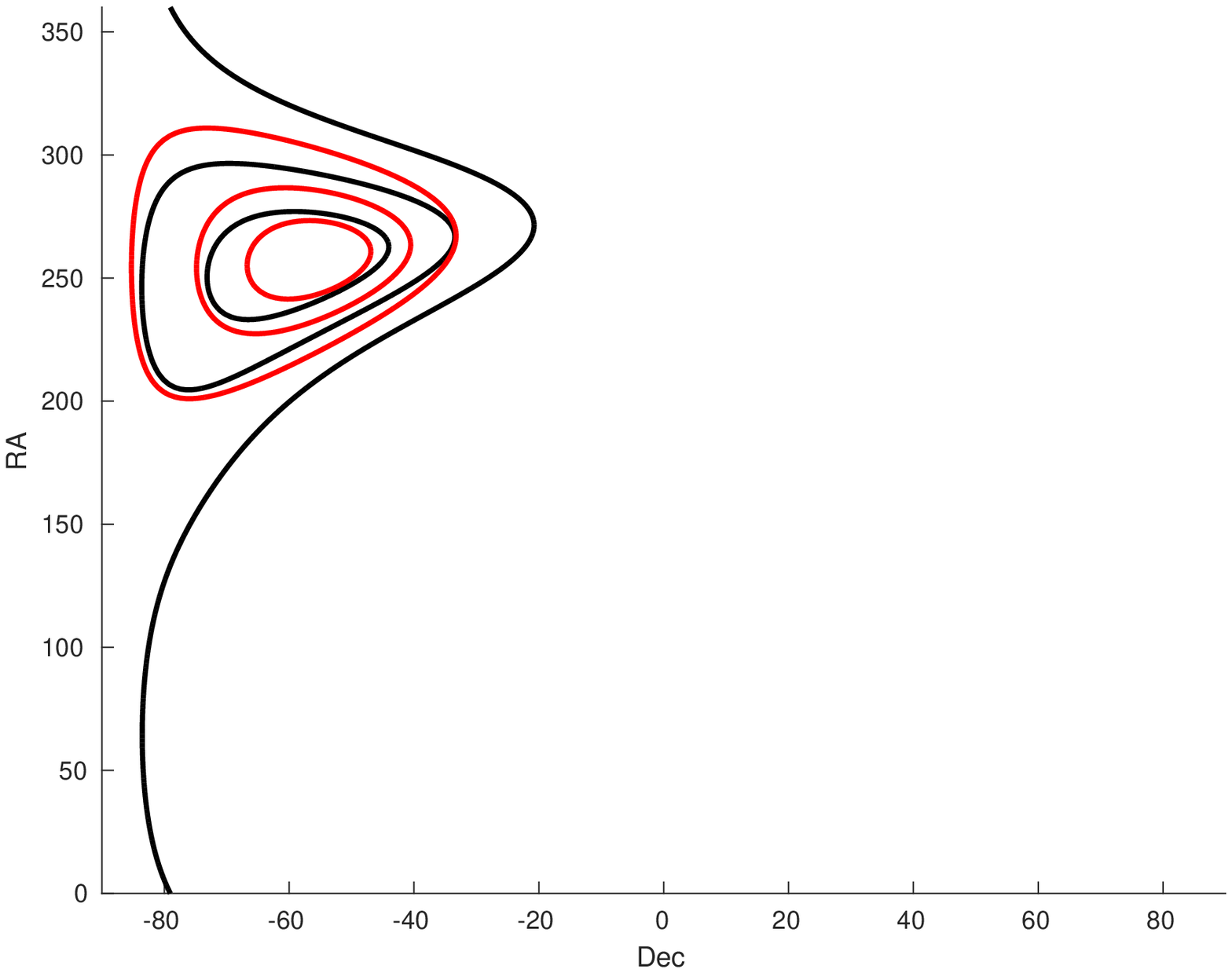}\\
\includegraphics[width=3in,keepaspectratio]{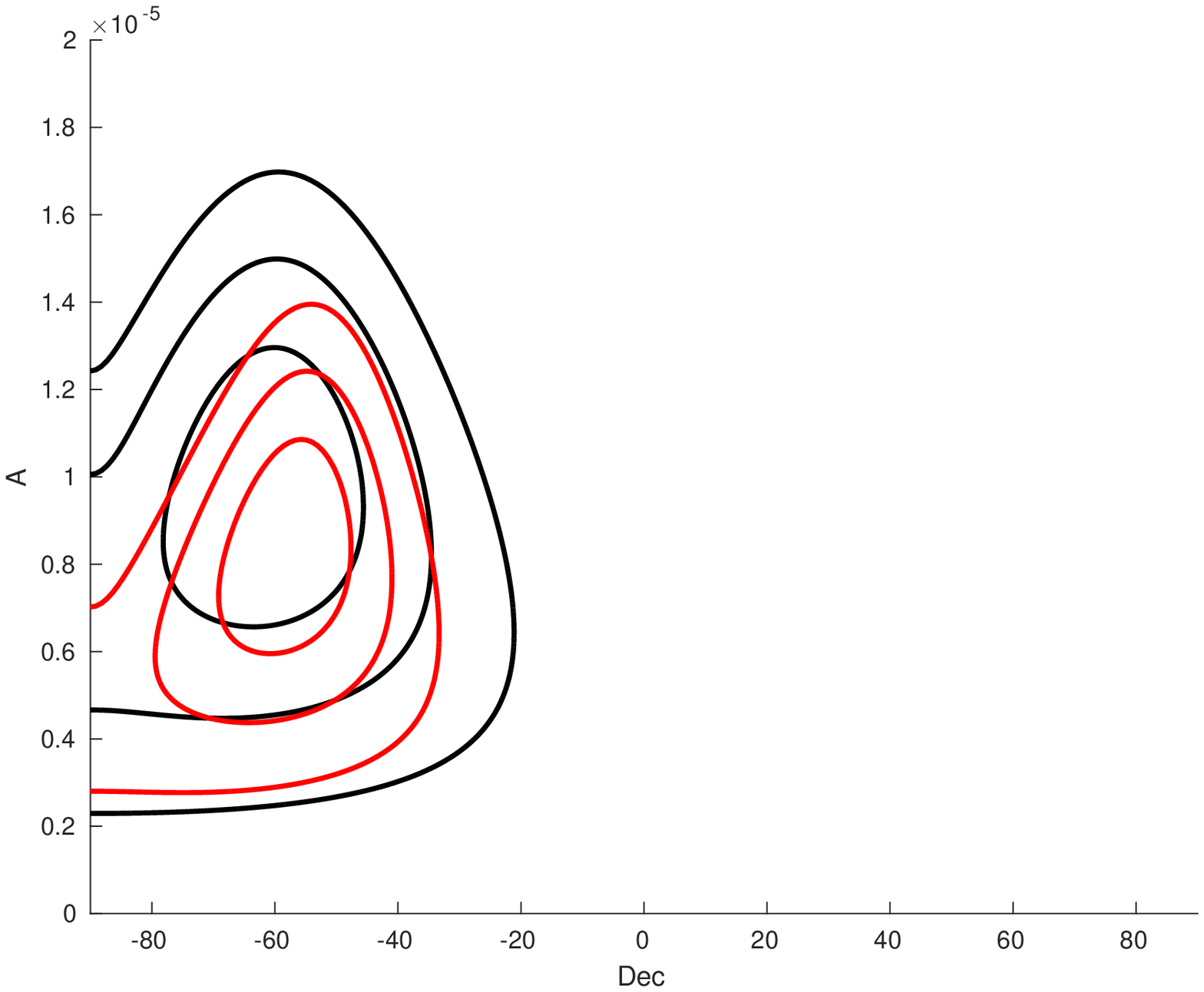}\\
\includegraphics[width=3in,keepaspectratio]{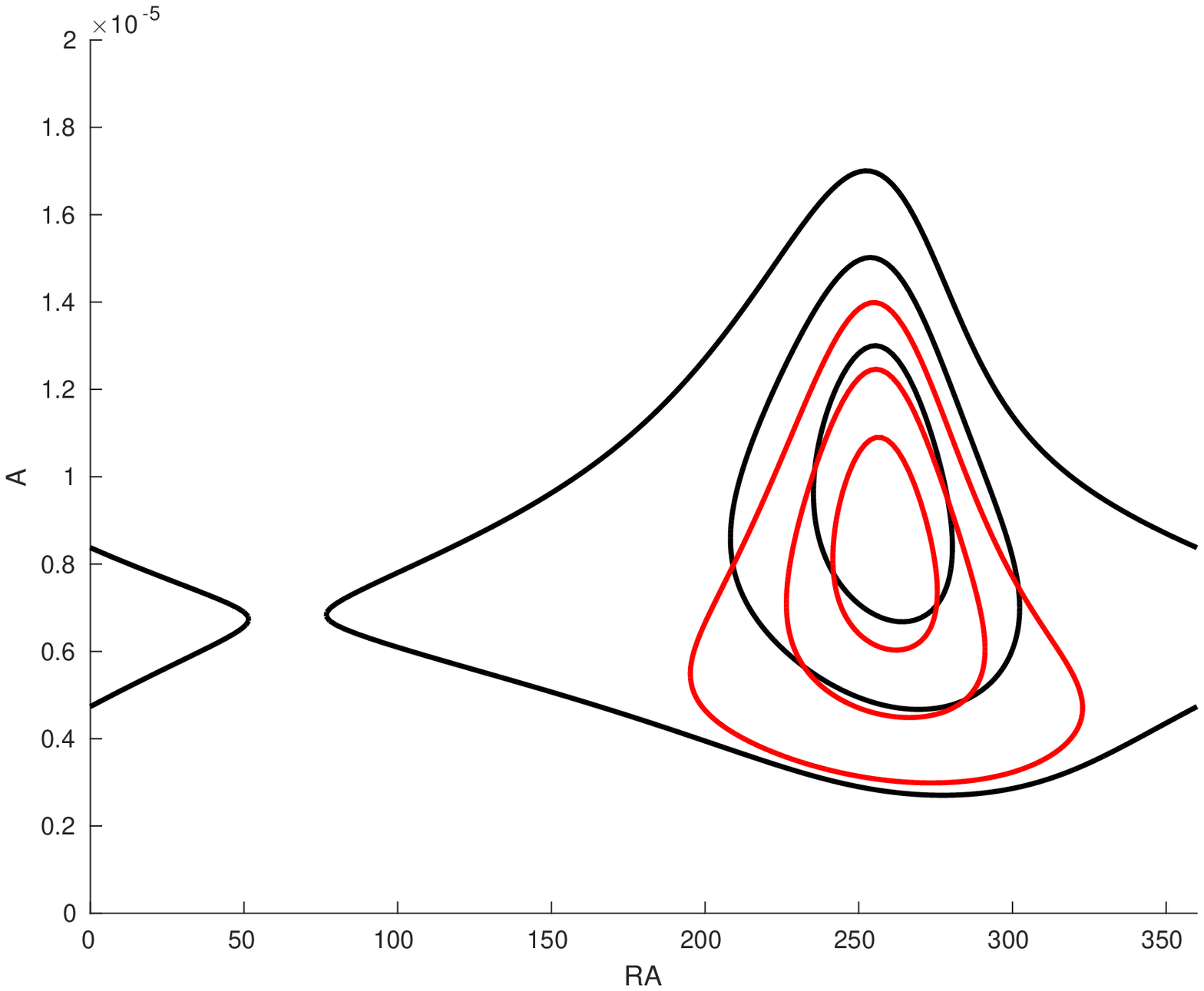}
\end{center}
\caption{2D likelihood contours for the Amplitude and coordinates of maximal variation (Right Ascension and Declination), with the remaining parameter marginalized, for a pure spatial dipole variation of $\alpha$, see Eq. (\protect\ref{puredipole}). The black contours correspond to the data of Webb {\it et al.} \cite{Webb}, while in the red ones that data is combined with the one presented in Table \protect\ref{tablealpha}. One, two and three sigma confidence levels are displayed in all cases.}
\label{fig1}
\end{figure}
\begin{figure}
\begin{center}
\includegraphics[width=3in,keepaspectratio]{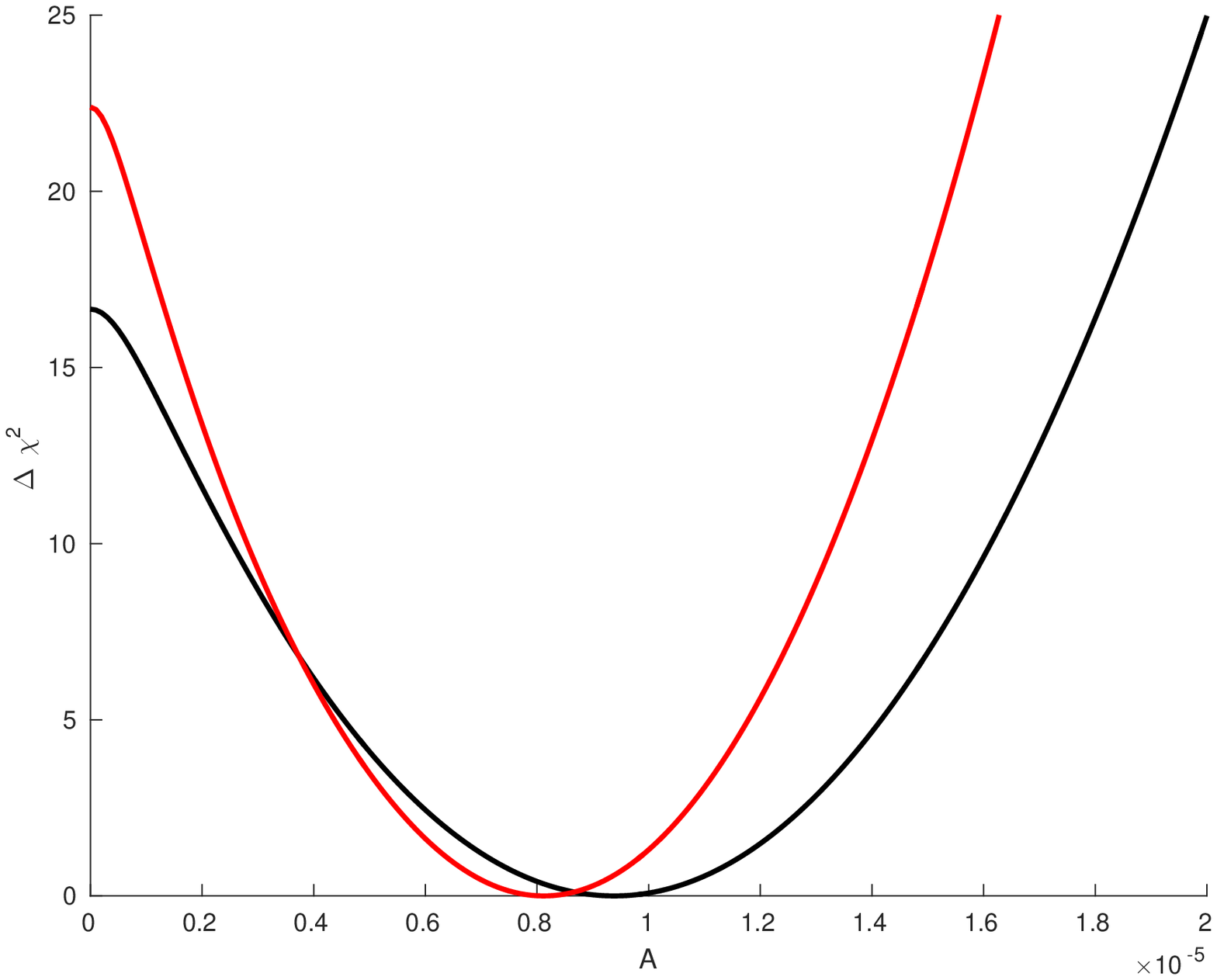}\\
\includegraphics[width=3in,keepaspectratio]{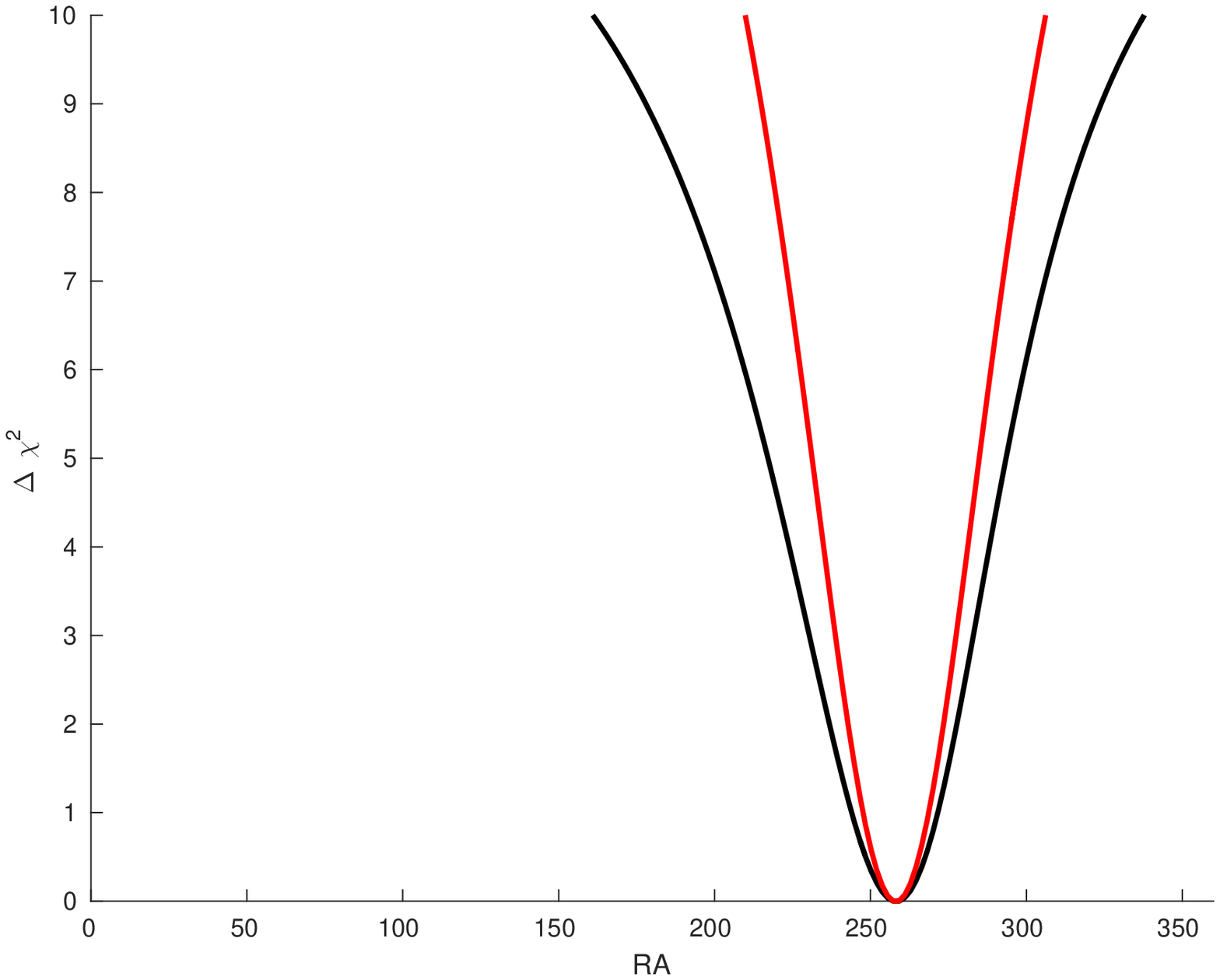}\\
\includegraphics[width=3in,keepaspectratio]{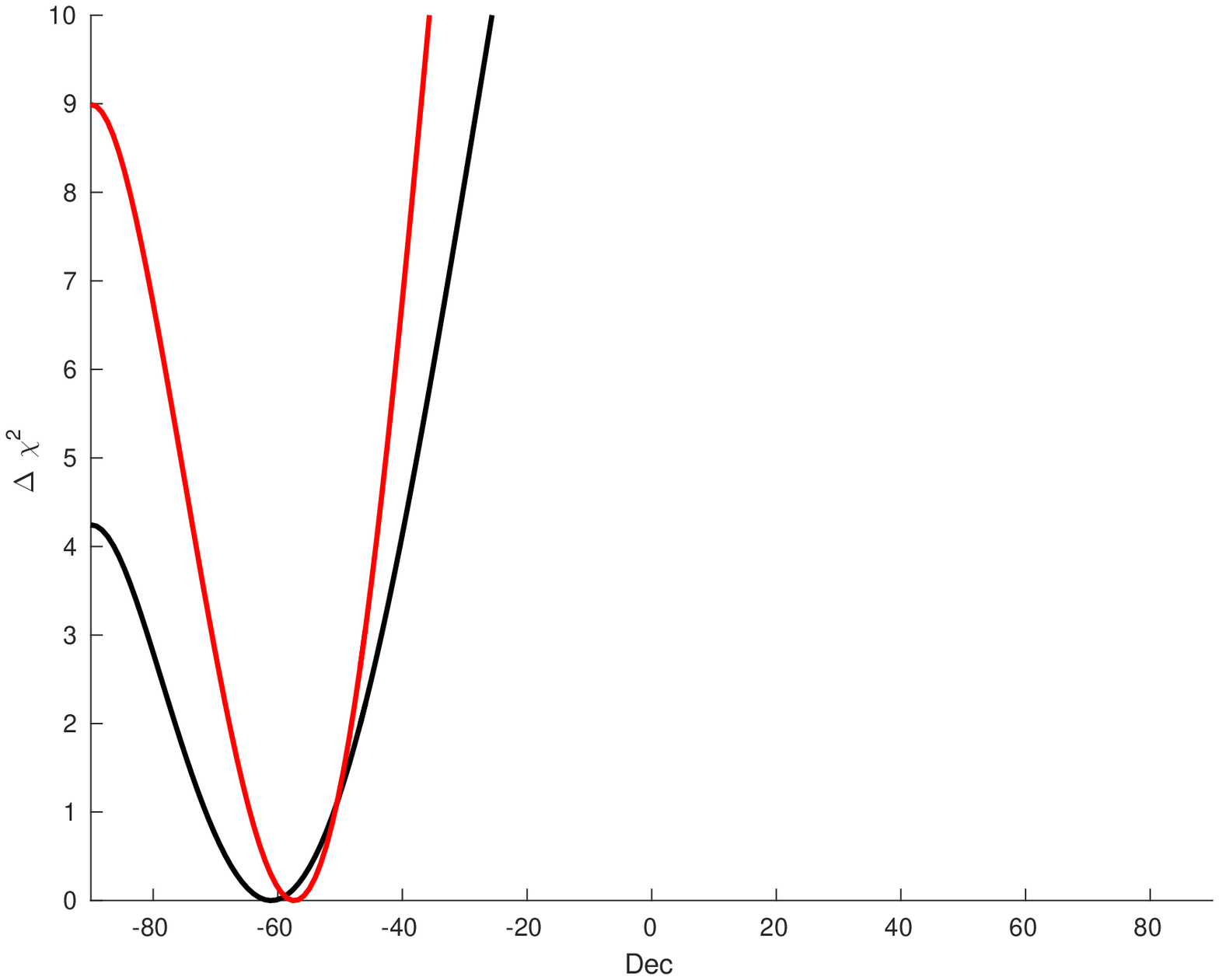}
\end{center}
\caption{1D likelihood for the Amplitude and coordinates of maximal variation (Right Ascension and declination), with the other parameter marginalized, for a pure spatial dipole variation of $\alpha$, see Eq. (\protect\ref{puredipole}). The black contours correspond to the data of Webb {\it et al.} \cite{Webb}, while in the red ones that data is combined with the one presented in Table \protect\ref{tablealpha}. The $\Delta\chi^2=\chi^2-\chi^2_{\rm min}$ is displayed in all cases.}
\label{fig2}
\end{figure}
\begin{table}
\begin{center}
\begin{tabular}{|c|c|c|c|}
\hline
Dataset \& c.l. & Amplitude ($ppm$) & Right Ascension ($h$) & Declination (${}^\circ$) \\
\hline
Webb {\it et al.} ($68.3\%$) & $9.4\pm2.2$ & $17.2\pm1.0$ & $-61\pm10$ \\
Webb {\it et al.} ($99.7\%$) & $9.4\pm6.4$ & $17.2^{+4.4}_{-5.3}$ & $<-28$ \\
\hline
All data ($68.3\%$) & $8.1\pm1.7$ & $17.2\pm0.7$ & $-58\pm7$ \\
All data ($99.7\%$) & $8.1\pm5.0$ & $17.2\pm2.9$ & $<-37$ \\
\hline
\end{tabular}
\caption{\label{tablepure}One- and three-sigma constraints on the Amplitude and coordinates of maximal variation (Right Ascension and declination) for a pure spatial dipole variation of $\alpha$. The 'all data' case corresponds to using the data of Webb {\it et al.} \cite{Webb} together with the 10 individual measurements presented in Table \protect\ref{tablealpha}. These results are also graphically displayed in Figure \protect\ref{fig2}.}
\end{center}
\end{table}

Figures \ref{fig1} and \ref{fig2} and Table \ref{tablepure} summarize the results of our analysis for the case of the pure spatial dipole. For the Webb {\it et al.} data alone we confirm the results of previous analyses. However, the addition of the more recent measurements has a significant impact on the results. While the statistical preference for a non-zero amplitude remains above the four-sigma level, the most likely value (and the corresponding uncertainty) for this amplitude decreases considerably, from $9.4$ to $8.1$ ppm. On the other hand the preferred direction of the north pole does not change significantly, but the corresponding uncertainties are reduced by about thirty percent in each coordinate.

\begin{figure}
\begin{center}
\includegraphics[width=3in,keepaspectratio]{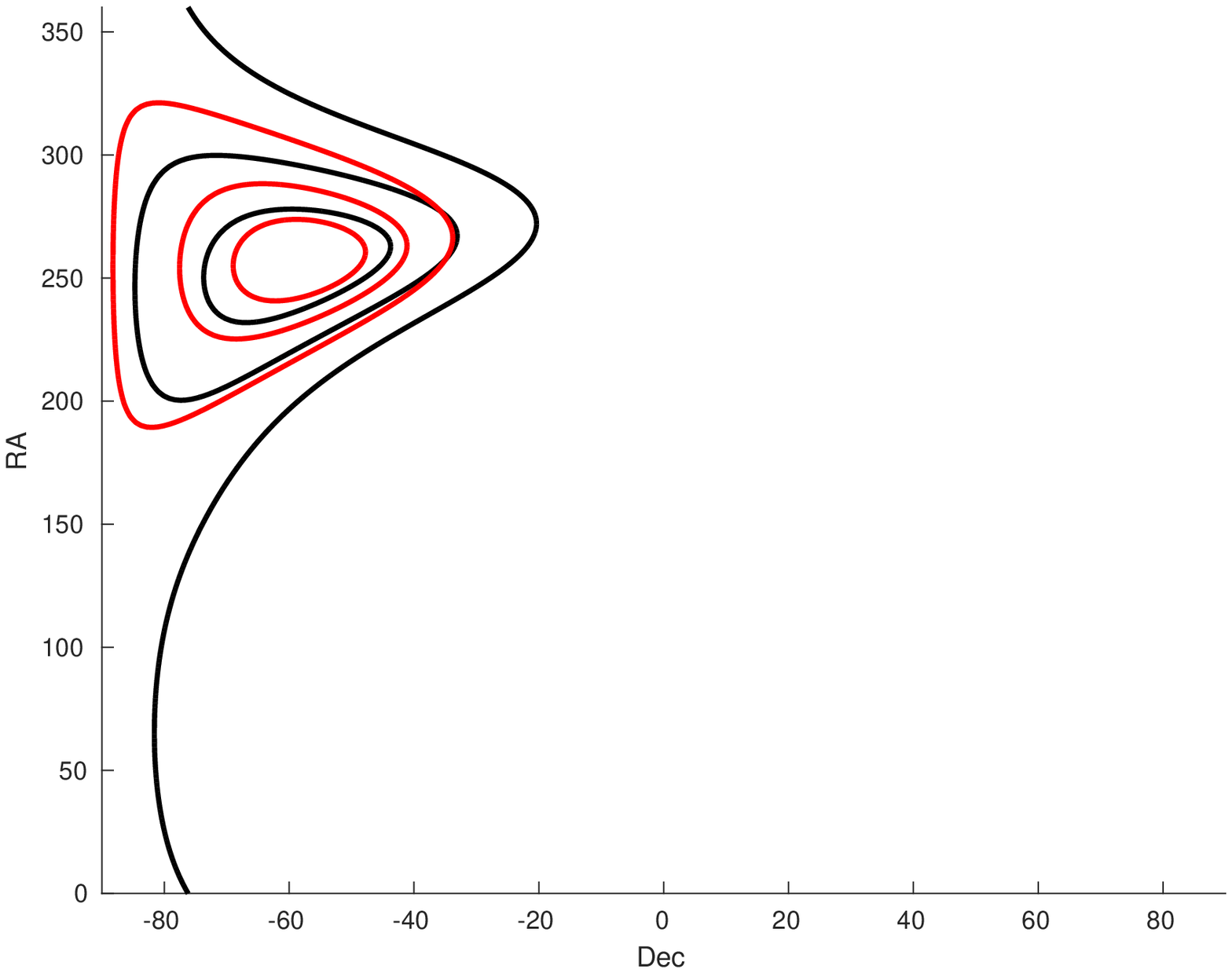}\\
\includegraphics[width=3in,keepaspectratio]{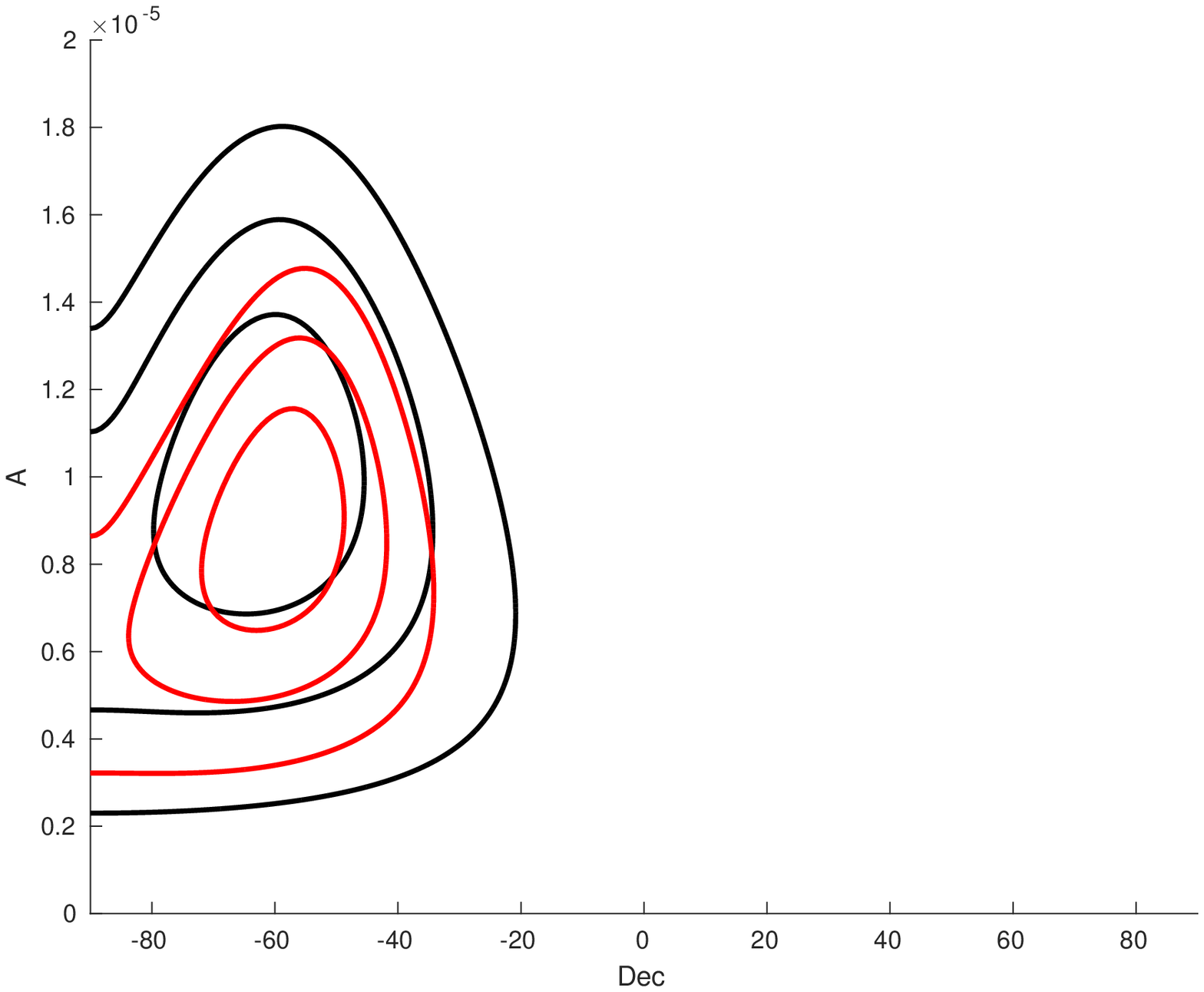}\\
\includegraphics[width=3in,keepaspectratio]{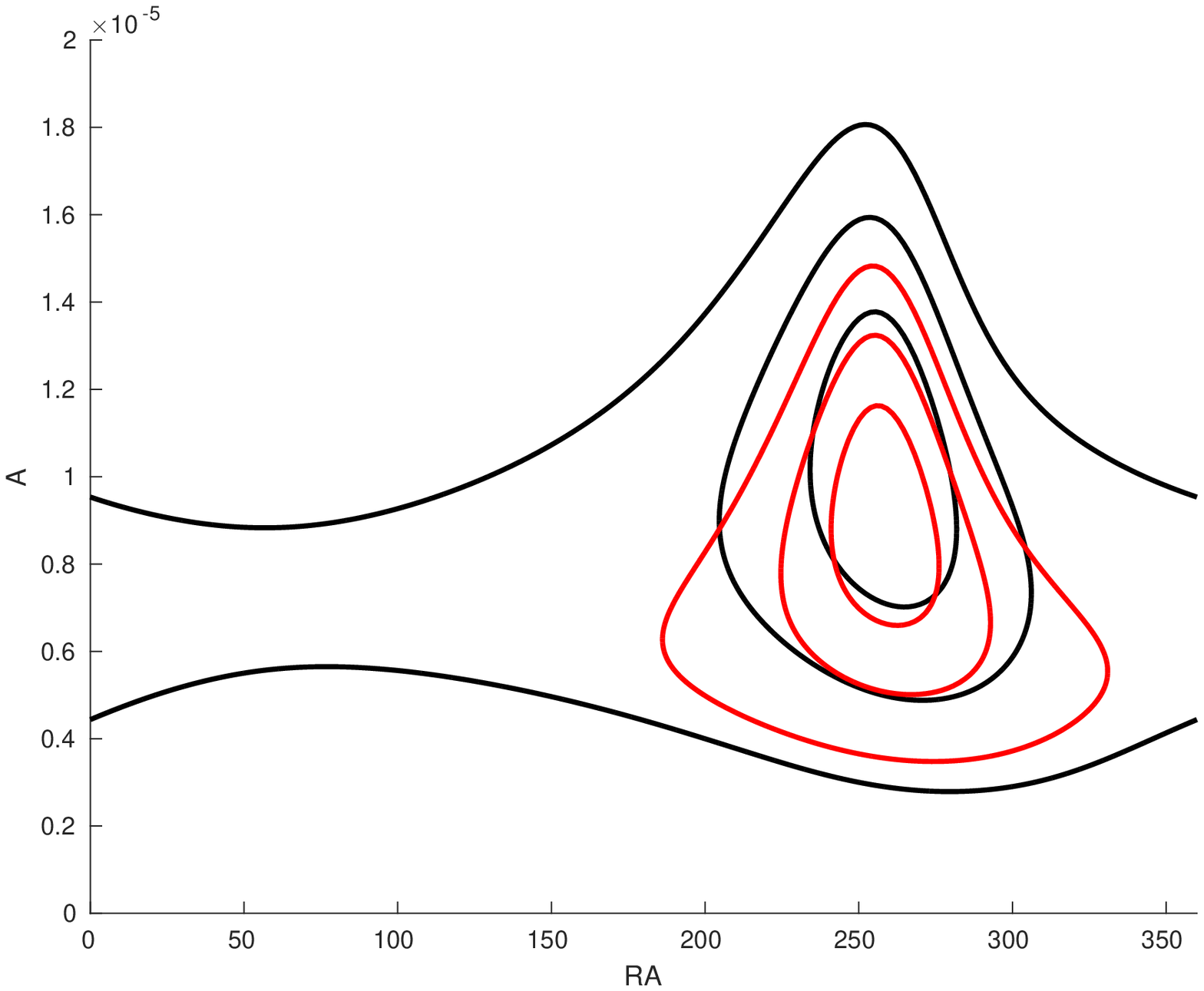}
\end{center}
\caption{2D likelihood contours for the Amplitude and coordinates of maximal variation (Right Ascension and Declination), with the remaining parameter marginalized, for a redshift-dependent dipole variation of $\alpha$, see Eq. (\protect\ref{redshiftdipole}). The black contours correspond to the data of Webb {\it et al.} \cite{Webb}, while in the red ones that data is combined with the one presented in Table \protect\ref{tablealpha}. One, two and three sigma confidence levels are displayed in all cases.}
\label{fig3}
\end{figure}
\begin{figure}
\begin{center}
\includegraphics[width=3in,keepaspectratio]{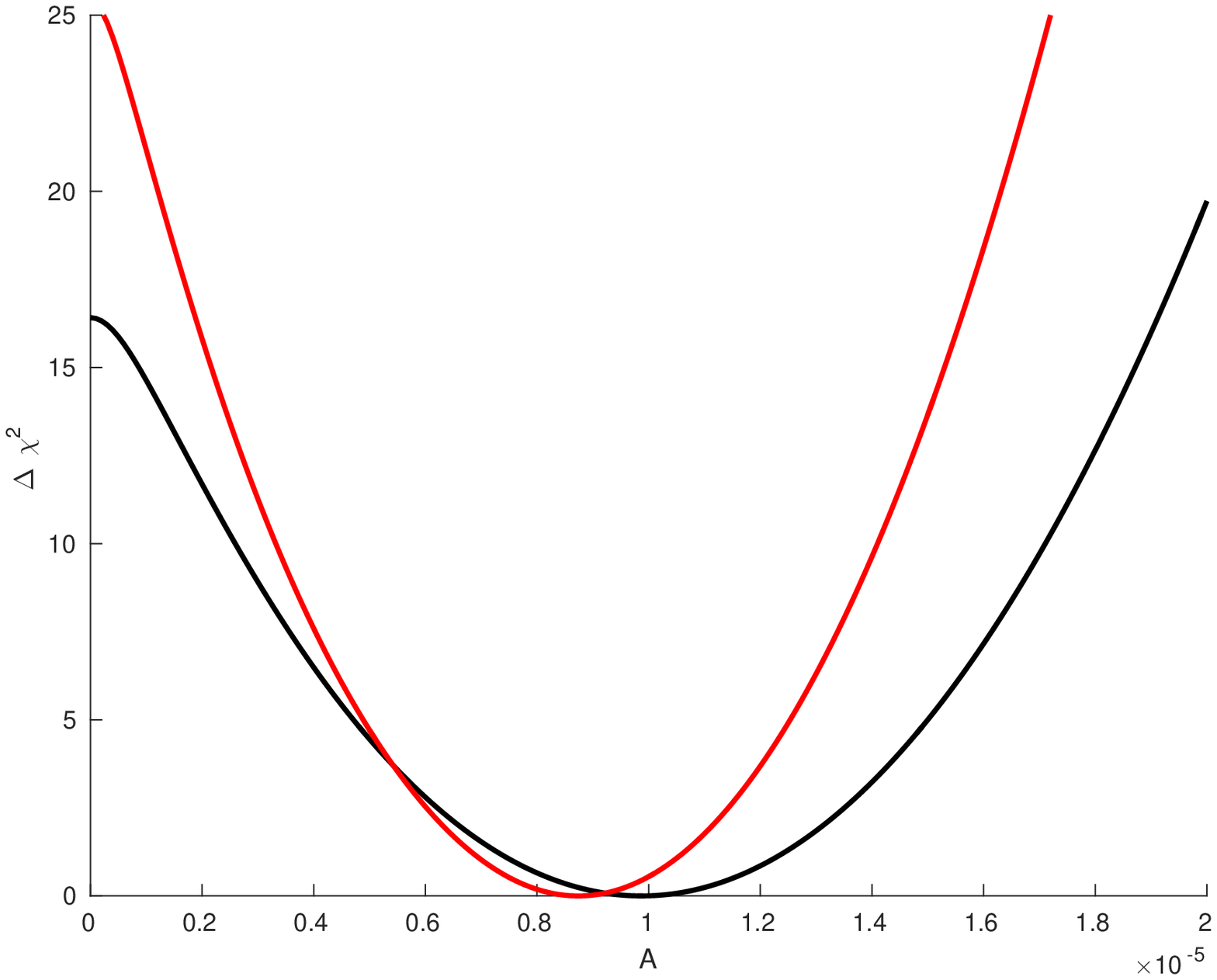}\\
\includegraphics[width=3in,keepaspectratio]{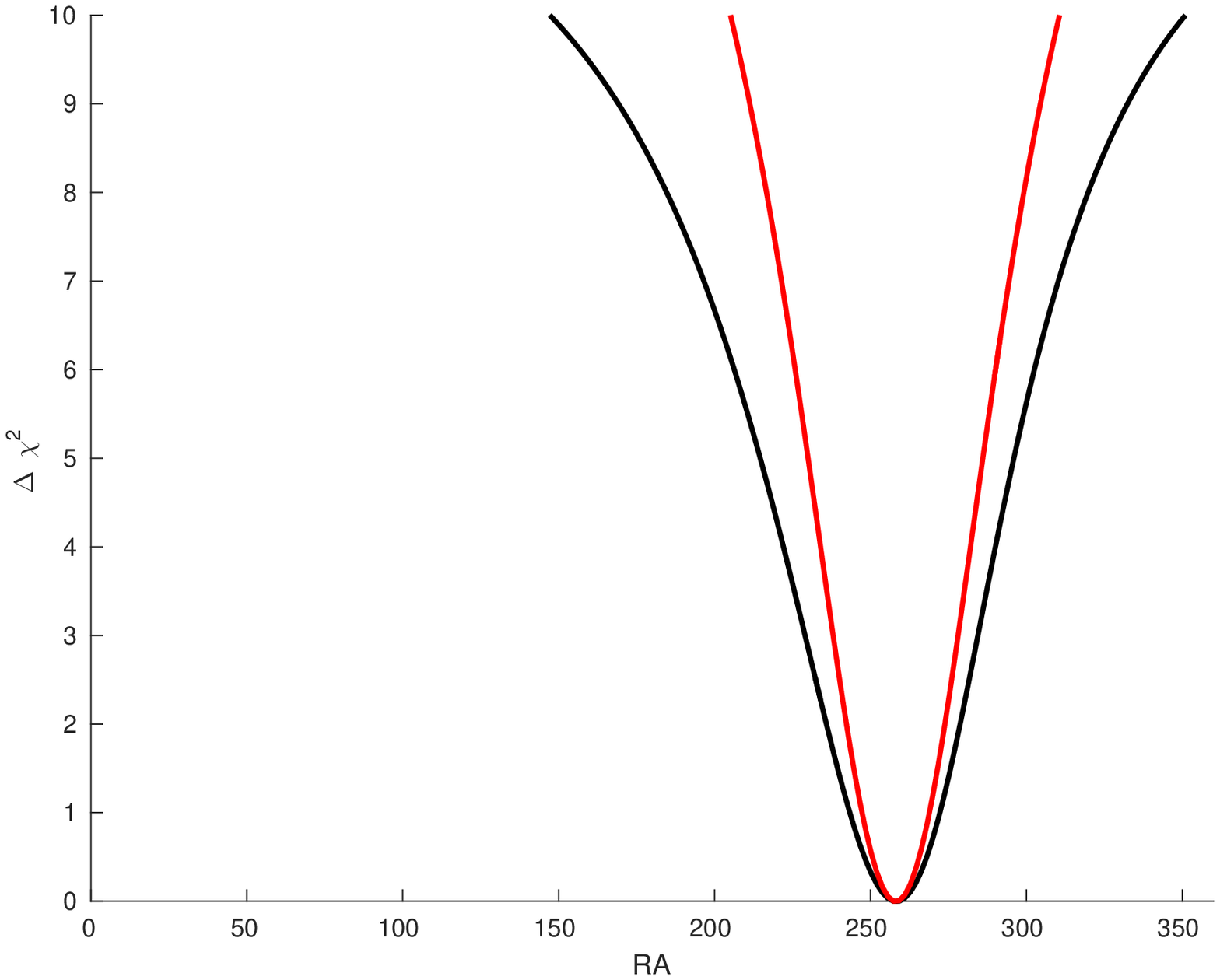}\\
\includegraphics[width=3in,keepaspectratio]{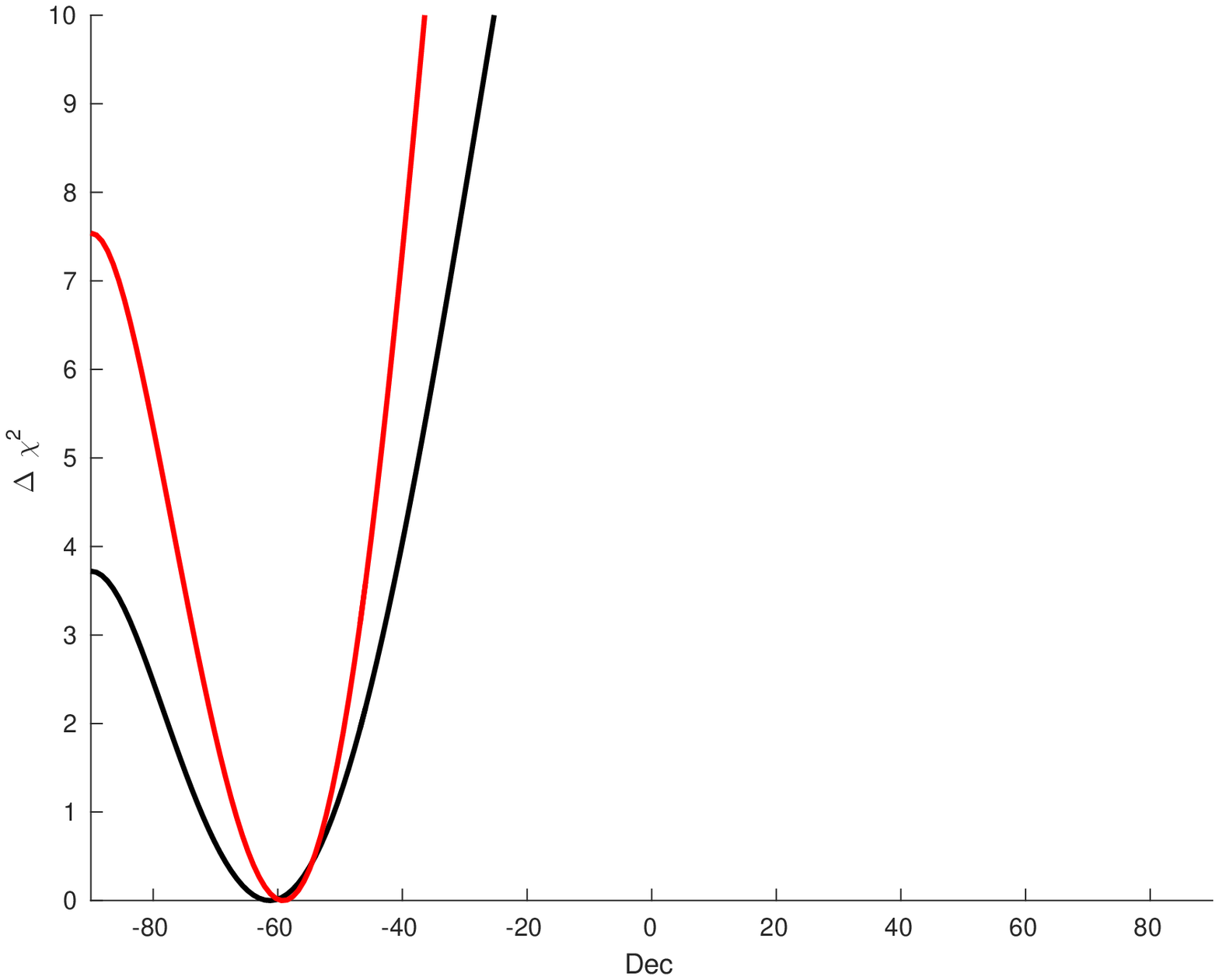}
\end{center}
\caption{1D likelihood for the Amplitude and coordinates of maximal variation (Right Ascension and declination), with the other parameter marginalized, for a redshift-dependent dipole variation of $\alpha$, see Eq. (\protect\ref{redshiftdipole}). The black contours correspond to the data of Webb {\it et al.} \cite{Webb}, while in the red ones that data is combined with the one presented in Table \protect\ref{tablealpha}. The $\Delta\chi^2=\chi^2-\chi^2_{\rm min}$ is displayed in all cases.}
\label{fig4}
\end{figure}
\begin{table}
\begin{center}
\begin{tabular}{|c|c|c|c|}
\hline
Dataset \& c.l. & Amplitude ($ppm$) & Right Ascension ($h$) & Declination (${}^\circ$) \\
\hline
Webb {\it et al.} ($68.3\%$) & $9.9\pm2.3$ & $17.2\pm1.0$ & $-61\pm11$ \\
Webb {\it et al.} ($99.7\%$) & $9.9\pm6.9$ & $17.2^{+5.0}_{-5.9}$ & $<-27$ \\
\hline
All data ($68.3\%$) & $8.7\pm1.7$ & $17.2\pm0.7$ & $-59\pm8$ \\
All data ($99.7\%$) & $8.7\pm5.1$ & $17.2\pm3.1$ & $<-38$ \\
\hline
\end{tabular}
\caption{\label{tableredshift}Same as Table \protect\ref{tablepure}, but for a dipolar variation with an additional redshift dependence, as given by Eq. (\protect\ref{redshiftdipole}). These results are also graphically displayed in Figure \protect\ref{fig4}.}
\end{center}
\end{table}

Figures \ref{fig3} and \ref{fig4} and Table \ref{tableredshift} contain analogous results for the redshift-dependent dipole. Again the statistical preference for a non-zero dipole is at more than four standard deviations, in this case with a slightly larger value of the preferred amplitude. The uncertainties in all three fitted parameters also increase slightly, as compared to the pure spatial dipole case. In any case we find, in agreement with previous works, that current data cannot strongly discriminate between the two classes of models.

\section{Outlook}

We have revisited recent indications of spatial variations of the fine-structure constant, $\alpha$, by considering the impact of the current set of dedicated measurements listed in Table \ref{tablealpha} on this analysis. While this dataset is currently still small, it has already been shown that it plays a significant role in obtaining constraints on dark energy and Weak Equivalence Principle violations \cite{Pinho}. Here we have confirmed that they also have a noticeable impact on constraints on spatial variations, thereby updating the original analysis of Webb {\it et al.}

Our analysis shows that a dipolar variation is still a good fit to the combined dataset, with the statistical preference for a non-zero amplitude remaining above the four-sigma level. However the addition of the new data reduces the best-fit amplitude as well as its uncertainty. The direction on the sky of the north pole of such a dipole remains almost unchanged, but its uncertainty is reduced by about thirty percent in each coordinate. Even with this additional data one can't yet statistically discriminate between a pure spatial dipole and one with an additional redshift dependence.

Naturally the key concern regarding these measurements is the possible presence of hidden systematics \cite{Whitmore}. Additional measurements from the ongoing UVES Large Program should shed further light on this subject. The dawn of a new generations of high-resolution ultra-stable spectrographs, of which ESPRESSO is the first example \cite{ESPRESSO}, will be a key development, allowing measurements not only with significantly smaller statistical uncertainties but also with a much better control over possible systematics. A roadmap for this field can be found in \cite{grg}.

\section*{Acknowledgements}
This work was done in the context of project PTDC/FIS/111725/2009 (FCT, Portugal). CJM is also supported by an FCT Research Professorship, contract reference IF/00064/2012, funded by FCT/MCTES (Portugal) and POPH/FSE. We are grateful to Ana Catarina Leite, Matteo Martinelli and Paolo Molaro for useful discussions on this topic.

\bibliographystyle{model1-num-names}
\bibliography{dipole}

\begin{thebibliography}{20}
\expandafter\ifx\csname natexlab\endcsname\relax\def\natexlab#1{#1}\fi
\providecommand{\bibinfo}[2]{#2}
\ifx\xfnm\relax \def\xfnm[#1]{\unskip,\space#1}\fi
\bibitem[{Uzan(2011)}]{Uzan}
\bibinfo{author}{J.-P. Uzan},
\newblock \bibinfo{title}{{Varying Constants, Gravitation and Cosmology}},
\newblock \bibinfo{journal}{Living Rev. Rel.} \bibinfo{volume}{14}
  (\bibinfo{year}{2011}) \bibinfo{pages}{2}.
\bibitem[{Martins(2015)}]{grg}
\bibinfo{author}{C.~J. A.~P. Martins},
\newblock \bibinfo{title}{{Fundamental cosmology in the E-ELT era: The status
  and future role of tests of fundamental coupling stability}},
\newblock \bibinfo{journal}{Gen.Rel.Grav.} \bibinfo{volume}{47}
  (\bibinfo{year}{2015}) \bibinfo{pages}{1843}.
\bibitem[{Webb et~al.(2011)Webb, King, Murphy, Flambaum, Carswell
  et~al.}]{Webb}
\bibinfo{author}{J.~Webb}, \bibinfo{author}{J.~King},
  \bibinfo{author}{M.~Murphy}, \bibinfo{author}{V.~Flambaum},
  \bibinfo{author}{R.~Carswell}, et~al.,
\newblock \bibinfo{title}{{Indications of a spatial variation of the fine
  structure constant}},
\newblock \bibinfo{journal}{Phys.Rev.Lett.} \bibinfo{volume}{107}
  (\bibinfo{year}{2011}) \bibinfo{pages}{191101}.
\bibitem[{King et~al.(2012)King, Webb, Murphy, Flambaum, Carswell, Bainbridge,
  Wilczynska, and Koch}]{King}
\bibinfo{author}{J.~A. King}, \bibinfo{author}{J.~K. Webb},
  \bibinfo{author}{M.~T. Murphy}, \bibinfo{author}{V.~V. Flambaum},
  \bibinfo{author}{R.~F. Carswell}, \bibinfo{author}{M.~B. Bainbridge},
  \bibinfo{author}{M.~R. Wilczynska}, \bibinfo{author}{F.~E. Koch},
\newblock \bibinfo{title}{{Spatial variation in the fine-structure constant --
  new results from VLT/UVES}},
\newblock \bibinfo{journal}{Mon. Not. Roy. Astron. Soc.} \bibinfo{volume}{422}
  (\bibinfo{year}{2012}) \bibinfo{pages}{3370--3413}.
\bibitem[{Berengut et~al.(2011)Berengut, Flambaum, King, Curran, and
  Webb}]{tests1}
\bibinfo{author}{J.~C. Berengut}, \bibinfo{author}{V.~V. Flambaum},
  \bibinfo{author}{J.~A. King}, \bibinfo{author}{S.~J. Curran},
  \bibinfo{author}{J.~K. Webb},
\newblock \bibinfo{title}{{Is there further evidence for spatial variation of
  fundamental constants?}},
\newblock \bibinfo{journal}{Phys. Rev.} \bibinfo{volume}{D83}
  (\bibinfo{year}{2011}) \bibinfo{pages}{123506}.
\bibitem[{Berengut et~al.(2012)Berengut, Kava, and Flambaum}]{tests2}
\bibinfo{author}{J.~C. Berengut}, \bibinfo{author}{E.~M. Kava},
  \bibinfo{author}{V.~V. Flambaum},
\newblock \bibinfo{title}{{A reanalysis of quasar absorption spectra results
  suggesting a spatial gradient in values of the fine-structure constant}},
\newblock \bibinfo{journal}{Astron. Astrophys.} \bibinfo{volume}{542}
  (\bibinfo{year}{2012}) \bibinfo{pages}{A118}.
\bibitem[{Mariano and Perivolaropoulos(2012)}]{tests3}
\bibinfo{author}{A.~Mariano}, \bibinfo{author}{L.~Perivolaropoulos},
\newblock \bibinfo{title}{{Is there correlation between Fine Structure and Dark
  Energy Cosmic Dipoles?}},
\newblock \bibinfo{journal}{Phys. Rev.} \bibinfo{volume}{D86}
  (\bibinfo{year}{2012}) \bibinfo{pages}{083517}.
\bibitem[{Kraiselburd et~al.(2013)Kraiselburd, Landau, and Simeone}]{tests4}
\bibinfo{author}{L.~Kraiselburd}, \bibinfo{author}{S.~J. Landau},
  \bibinfo{author}{C.~Simeone},
\newblock \bibinfo{title}{{Variation of the fine-structure constant: an update
  of statistical analyses with recent data}},
\newblock \bibinfo{journal}{Astron. Astrophys.} \bibinfo{volume}{557}
  (\bibinfo{year}{2013}) \bibinfo{pages}{A36}.
\bibitem[{Molaro et~al.(2013)Molaro, Centurion, Whitmore, Evans, Murphy
  et~al.}]{LP1}
\bibinfo{author}{P.~Molaro}, \bibinfo{author}{M.~Centurion},
  \bibinfo{author}{J.~Whitmore}, \bibinfo{author}{T.~Evans},
  \bibinfo{author}{M.~Murphy}, et~al.,
\newblock \bibinfo{title}{{The UVES Large Program for Testing Fundamental
  Physics: I Bounds on a change in alpha towards quasar HE 2217-2818}},
\newblock \bibinfo{journal}{Astron.Astrophys.} \bibinfo{volume}{555}
  (\bibinfo{year}{2013}) \bibinfo{pages}{A68}.
\bibitem[{{Evans} et~al.(2014){Evans}, {Murphy}, {Whitmore}, {Misawa},
  {Centurion}, {D'Odorico}, {Lopez}, {Martins}, {Molaro}, {Petitjean},
  {Rahmani}, {Srianand}, and {Wendt}}]{LP3}
\bibinfo{author}{T.~M. {Evans}}, \bibinfo{author}{M.~T. {Murphy}},
  \bibinfo{author}{J.~B. {Whitmore}}, \bibinfo{author}{T.~{Misawa}},
  \bibinfo{author}{M.~{Centurion}}, \bibinfo{author}{S.~{D'Odorico}},
  \bibinfo{author}{S.~{Lopez}}, \bibinfo{author}{C.~J.~A.~P. {Martins}},
  \bibinfo{author}{P.~{Molaro}}, \bibinfo{author}{P.~{Petitjean}},
  \bibinfo{author}{H.~{Rahmani}}, \bibinfo{author}{R.~{Srianand}},
  \bibinfo{author}{M.~{Wendt}},
\newblock \bibinfo{title}{{The UVES Large Program for testing fundamental
  physics - III. Constraints on the fine-structure constant from three
  telescopes}},
\newblock \bibinfo{journal}{M.N.R.A.S.} \bibinfo{volume}{445}
  (\bibinfo{year}{2014}) \bibinfo{pages}{128--150}.
\bibitem[{Whitmore and Murphy(2015)}]{Whitmore}
\bibinfo{author}{J.~B. Whitmore}, \bibinfo{author}{M.~T. Murphy},
\newblock \bibinfo{title}{{Impact of instrumental systematic errors on
  fine-structure constant measurements with quasar spectra}},
\newblock \bibinfo{journal}{Mon. Not. Roy. Astron. Soc.} \bibinfo{volume}{447}
  (\bibinfo{year}{2015}) \bibinfo{pages}{446--462}.
\bibitem[{Ferreira et~al.(2014)Ferreira, Frigola, Martins, Monteiro, and
  Sol\`a}]{Ferreira}
\bibinfo{author}{M.~C. Ferreira}, \bibinfo{author}{O.~Frigola},
  \bibinfo{author}{C.~J. A.~P. Martins}, \bibinfo{author}{A.~M. R. V.~L.
  Monteiro}, \bibinfo{author}{J.~Sol\`a},
\newblock \bibinfo{title}{{Consistency tests of the stability of fundamental
  couplings and unification scenarios}},
\newblock \bibinfo{journal}{Phys. Rev.} \bibinfo{volume}{D89}
  (\bibinfo{year}{2014}) \bibinfo{pages}{083011}.
\bibitem[{Songaila and Cowie(2014)}]{Songaila}
\bibinfo{author}{A.~Songaila}, \bibinfo{author}{L.~Cowie},
\newblock \bibinfo{title}{{Constraining the Variation of the Fine Structure
  Constant with Observations of Narrow Quasar Absorption Lines}},
\newblock \bibinfo{journal}{Astrophys.J.} \bibinfo{volume}{793}
  (\bibinfo{year}{2014}) \bibinfo{pages}{103}.
\bibitem[{Molaro et~al.(2008)Molaro, Reimers, Agafonova, and
  Levshakov}]{alphaMolaro}
\bibinfo{author}{P.~Molaro}, \bibinfo{author}{D.~Reimers},
  \bibinfo{author}{I.~I. Agafonova}, \bibinfo{author}{S.~A. Levshakov},
\newblock \bibinfo{title}{{Bounds on the fine structure constant variability
  from FeII absorption lines in QSO spectra}},
\newblock \bibinfo{journal}{Eur.Phys.J.ST} \bibinfo{volume}{163}
  (\bibinfo{year}{2008}) \bibinfo{pages}{173--189}.
\bibitem[{{Chand} et~al.(2006){Chand}, {Srianand}, {Petitjean}, {Aracil},
  {Quast}, and {Reimers}}]{alphaChand}
\bibinfo{author}{H.~{Chand}}, \bibinfo{author}{R.~{Srianand}},
  \bibinfo{author}{P.~{Petitjean}}, \bibinfo{author}{B.~{Aracil}},
  \bibinfo{author}{R.~{Quast}}, \bibinfo{author}{D.~{Reimers}},
\newblock \bibinfo{title}{{Variation of the fine-structure constant: very high
  resolution spectrum of QSO HE 0515-4414}},
\newblock \bibinfo{journal}{Astron.Astrophys.} \bibinfo{volume}{451}
  (\bibinfo{year}{2006}) \bibinfo{pages}{45--56}.
\bibitem[{{Agafonova} et~al.(2011){Agafonova}, {Molaro}, {Levshakov}, and
  {Hou}}]{alphaAgafonova}
\bibinfo{author}{I.~I. {Agafonova}}, \bibinfo{author}{P.~{Molaro}},
  \bibinfo{author}{S.~A. {Levshakov}}, \bibinfo{author}{J.~L. {Hou}},
\newblock \bibinfo{title}{{First measurement of Mg isotope abundances at high
  redshifts and accurate estimate of {$\Delta$}{$\alpha$}/{$\alpha$}}},
\newblock \bibinfo{journal}{Astron.Astrophys.} \bibinfo{volume}{529}
  (\bibinfo{year}{2011}) \bibinfo{pages}{A28}.
\bibitem[{Webb et~al.(2014)Webb, Wright, Koch, and Murphy}]{Monopole}
\bibinfo{author}{J.~K. Webb}, \bibinfo{author}{A.~Wright},
  \bibinfo{author}{F.~E. Koch}, \bibinfo{author}{M.~T. Murphy},
\newblock \bibinfo{title}{{Enhanced heavy magnesium isotopes in quasar
  absorption systems and varying alpha}},
\newblock \bibinfo{journal}{Mem. Soc. Ast. It.} \bibinfo{volume}{85}
  (\bibinfo{year}{2014}) \bibinfo{pages}{57--62}.
\bibitem[{Martins et~al.(2015{\natexlab{a}})Martins, Vielzeuf, Martinelli,
  Calabrese, and Pandolfi}]{Dilaton}
\bibinfo{author}{C.~J. A.~P. Martins}, \bibinfo{author}{P.~E. Vielzeuf},
  \bibinfo{author}{M.~Martinelli}, \bibinfo{author}{E.~Calabrese},
  \bibinfo{author}{S.~Pandolfi},
\newblock \bibinfo{title}{{Evolution of the fine-structure constant in runaway
  dilaton models}},
\newblock \bibinfo{journal}{Phys. Lett.} \bibinfo{volume}{B743}
  (\bibinfo{year}{2015}{\natexlab{a}}) \bibinfo{pages}{377--382}.
\bibitem[{Martins et~al.(2015{\natexlab{b}})Martins, Pinho, Alves, Pino, Rocha,
  and von Wietersheim}]{Pinho}
\bibinfo{author}{C.~J. A.~P. Martins}, \bibinfo{author}{A.~M.~M. Pinho},
  \bibinfo{author}{R.~F.~C. Alves}, \bibinfo{author}{M.~Pino},
  \bibinfo{author}{C.~I. S.~A. Rocha}, \bibinfo{author}{M.~von Wietersheim},
\newblock \bibinfo{title}{{Dark energy and Equivalence Principle constraints
  from astrophysical tests of the stability of the fine-structure constant}},
\newblock \bibinfo{journal}{JCAP} \bibinfo{volume}{1508}
  (\bibinfo{year}{2015}{\natexlab{b}}) \bibinfo{pages}{047}.
\bibitem[{{Pepe} et~al.(2013){Pepe}, {Cristiani}, {Rebolo}, {Santos}, {Dekker},
  {M{\'e}gevand}, {Zerbi}, {Cabral}, {Molaro}, {Di Marcantonio}, {Abreu},
  {Affolter}, {Aliverti}, {Allende Prieto}, {Amate}, {Avila}, {Baldini},
  {Bristow}, {Broeg}, {Cirami}, {Coelho}, {Conconi}, {Coretti}, {Cupani},
  {D'Odorico}, {De Caprio}, {Delabre}, {Dorn}, {Figueira}, {Fragoso},
  {Galeotta}, {Genolet}, {Gomes}, {Gonz{\'a}lez Hern{\'a}ndez}, {Hughes},
  {Iwert}, {Kerber}, {Landoni}, {Lizon}, {Lovis}, {Maire}, {Mannetta},
  {Martins}, {Monteiro}, {Oliveira}, {Poretti}, {Rasilla}, {Riva}, {Santana
  Tschudi}, {Santos}, {Sosnowska}, {Sousa}, {Span{\`o}}, {Tenegi}, {Toso},
  {Vanzella}, {Viel}, and {Zapatero Osorio}}]{ESPRESSO}
\bibinfo{author}{F.~{Pepe}}, \bibinfo{author}{S.~{Cristiani}},
  \bibinfo{author}{R.~{Rebolo}}, \bibinfo{author}{N.~C. {Santos}},
  \bibinfo{author}{H.~{Dekker}}, \bibinfo{author}{D.~{M{\'e}gevand}},
  \bibinfo{author}{F.~M. {Zerbi}}, \bibinfo{author}{A.~{Cabral}},
  \bibinfo{author}{P.~{Molaro}}, \bibinfo{author}{P.~{Di Marcantonio}},
  \bibinfo{author}{M.~{Abreu}}, \bibinfo{author}{M.~{Affolter}},
  \bibinfo{author}{M.~{Aliverti}}, \bibinfo{author}{C.~{Allende Prieto}},
  \bibinfo{author}{M.~{Amate}}, \bibinfo{author}{G.~{Avila}},
  \bibinfo{author}{V.~{Baldini}}, \bibinfo{author}{P.~{Bristow}},
  \bibinfo{author}{C.~{Broeg}}, \bibinfo{author}{R.~{Cirami}},
  \bibinfo{author}{J.~{Coelho}}, \bibinfo{author}{P.~{Conconi}},
  \bibinfo{author}{I.~{Coretti}}, \bibinfo{author}{G.~{Cupani}},
  \bibinfo{author}{V.~{D'Odorico}}, \bibinfo{author}{V.~{De Caprio}},
  \bibinfo{author}{B.~{Delabre}}, \bibinfo{author}{R.~{Dorn}},
  \bibinfo{author}{P.~{Figueira}}, \bibinfo{author}{A.~{Fragoso}},
  \bibinfo{author}{S.~{Galeotta}}, \bibinfo{author}{L.~{Genolet}},
  \bibinfo{author}{R.~{Gomes}}, \bibinfo{author}{J.~I. {Gonz{\'a}lez
  Hern{\'a}ndez}}, \bibinfo{author}{I.~{Hughes}}, \bibinfo{author}{O.~{Iwert}},
  \bibinfo{author}{F.~{Kerber}}, \bibinfo{author}{M.~{Landoni}},
  \bibinfo{author}{J.-L. {Lizon}}, \bibinfo{author}{C.~{Lovis}},
  \bibinfo{author}{C.~{Maire}}, \bibinfo{author}{M.~{Mannetta}},
  \bibinfo{author}{C.~{Martins}}, \bibinfo{author}{M.~A. {Monteiro}},
  \bibinfo{author}{A.~{Oliveira}}, \bibinfo{author}{E.~{Poretti}},
  \bibinfo{author}{J.~L. {Rasilla}}, \bibinfo{author}{M.~{Riva}},
  \bibinfo{author}{S.~{Santana Tschudi}}, \bibinfo{author}{P.~{Santos}},
  \bibinfo{author}{D.~{Sosnowska}}, \bibinfo{author}{S.~{Sousa}},
  \bibinfo{author}{P.~{Span{\`o}}}, \bibinfo{author}{F.~{Tenegi}},
  \bibinfo{author}{G.~{Toso}}, \bibinfo{author}{E.~{Vanzella}},
  \bibinfo{author}{M.~{Viel}}, \bibinfo{author}{M.~R. {Zapatero Osorio}},
\newblock \bibinfo{title}{{ESPRESSO---An Echelle SPectrograph for Rocky
  Exoplanets Search and Stable Spectroscopic Observations}},
\newblock \bibinfo{journal}{The Messenger} \bibinfo{volume}{153}
  (\bibinfo{year}{2013}) \bibinfo{pages}{6--16}.

\end{thebibliography}

\end{document}